\documentclass[12pt]{article}

\usepackage[applemac]{inputenc}
\usepackage[T1]{fontenc}
\usepackage{amsmath}
\usepackage{amsfonts}
\usepackage[english]{babel}

\usepackage{geometry}                
\usepackage[parfill]{parskip}    
\usepackage{graphicx}
\usepackage{amssymb}
\usepackage{epstopdf}
\usepackage[all]{xy}

\usepackage{amsthm} 

\theoremstyle{definition} 
 
\theoremstyle{axiom}

\theoremstyle{remark} 
 
\theoremstyle{plain}

\theoremstyle{plain}

\begin{document}

\title{On some (so-called) «paradoxes» of Quantum Physics}
\author{Daniel Parrochia}
\date{University of Lyon (France)}
\maketitle

\textbf{Abstract.}
This article is devoted to the study of which appears as the most famous paradoxes of Quantum Theory (Schrödinger's cat, EPR argument and Aspect's experiments, «delayed choice» experiments and retrocausality problems). Through these experiments, physics raises so fundamental questions that it borders, at the limit, with metaphysics. The present article supports the idea that the difficulties encountered, so puzzling they are, manifest only a transitional state of the evolution of physics, that can be expected to be outdated one day. In the meantime, caution is necessary to avoid the excesses that could lead to metaphysical considerations a little too premature.

\textbf{Key words.} Schrödinger's cat, EPR argument, Delayed-choice experiments, retrocausality, free-will theorem.

\section{Introduction}

Since 1905 and Max Planck's introduction of a discrete approach to physical experience, in order to solve the contradiction between the Raleigh-Jeans law and the law of Wien, and to find a unique explanation for electromagnetic radiation, Quantum Mechanics has not stopped posing problems to physicists. The very notion of irreducible quanta and the discontinuity of quantum phenomena remained rather mysterious in the 1920s. As one knows, Bohr and Heisenberg have developped the famous «Copenhagen interpretation» but the behavior of particles, reduced to observables and described by the infinite matrices of a Hilbert space, was still puzzling. Indeed, with the passage to the wave formalism, Planck had put his finger in a gear : there was not only an irreducible difficulty with the duality wave-corpuscule - valid, after the works of Louis de Broglie, for the matter as well as for the light -,  but physics, in general, was now going to encounter disturbing situations, very hard to interpret. If the notion of wave - once Schrödinger had associated an equation - had seemed to illuminate the situation for a moment, it had to be recognized that this wave, except in the case of a single particle, was not a usual wave. In collision experiments in particular, it was never diluted throughout the space. Moreover, for a system of $N$ particles, it propagated in a space of $3N$ dimensions. So Born went to give a purely probabilistic interpretation of waves, forcing Quantum Mechanics to give up a realistic description of particles, then bringing another mystery : the «reduction of the wave packet» at the moment of measurement. It became obvious that the wave function was therefore nothing conventional and even possessed paradoxical properties: on the one hand, it did not have the meaning of a real physical description, but on the other, it was not only a mathematical object, relative or contextual, which would have described only the knowledge we have of reality. The wave function (or its generalization, the state vector) was actually a subtle blend of the two. It was going to result in many paradoxical situations which we examine in the following, from the famous cat's Schrödinger thought experiment to delayed choice experiments, without forgetting, of course, the paradox EPR and its different solutions.

\section{Schrödinger's cat and others}

	The first paradox raises the question of the limit between the quantum phenomena and the classical (macroscopic) world of our daily experience. Everyone knows today the story of the Schrödinger's cat, which has elicited so much comment that it ended up irritating the famous physicist Stephen Hawking. However, remember the essence of the subject and its context.

As we know, the Schrödinger equation, governing the possible states of an elementary particle, is that of a linear wave function. This implies that for two possible states of the particle, the combination of the two states is also a possible state. On the other hand, the observation of the phenomenon causes, as we say, the «collapse» of the wave function, and therefore the reduction to a single state.

More precisely, it is the measure that disturbs the system - by what is called the Compton effect 
(diffusion of an electron by an incident photon whose wavelength increases) - and makes it bifurcate from a superimposed quantum state to a measured state. This state does not pre-exist the measure: it is the measure, apparently, that makes it happen.

The notion of «superposition of states», when it applies to a particle or an atom, does not raise a major problem in human mind. But, if one supposes a direct dependence between the state of these quantum realities and that of a macroscopic system - which suggests the linearity of the Schrödinger equation -, then it is another matter.

In 1935, in the context of Einstein's investigation about physical reality, Schrödinger - as Einstein himself praised it - was one of the few physicists to understand this perfectly puzzling implication of Quantum Mechanics. His famous thought experiment - known today as «Schrödinger's cat» - was precisely intended to reveal the problem raised and, as a result, to highlight the shortcomings of the Copenhagen interpretation, especially with respect to the problem of measurement.

Schrödinger thus supposes a thought experiment in which a cat is enclosed in a box with a device that kills the animal as soon as it detects the disintegration of an atom of a radioactive substance. For example, a Geiger radioactivity detector, connected to a switch, caused the fall of a hammer breaking a poison flask (Schrödinger proposed hydrocyanic acid) which can be enclosed in liquid form in a pressurized flask and vaporized, becoming a deadly gas once the bottle is broken.
	
	\begin{figure}[h] 
	   \centering
	      \vspace{-2\baselineskip}
	   \includegraphics[width=4in]{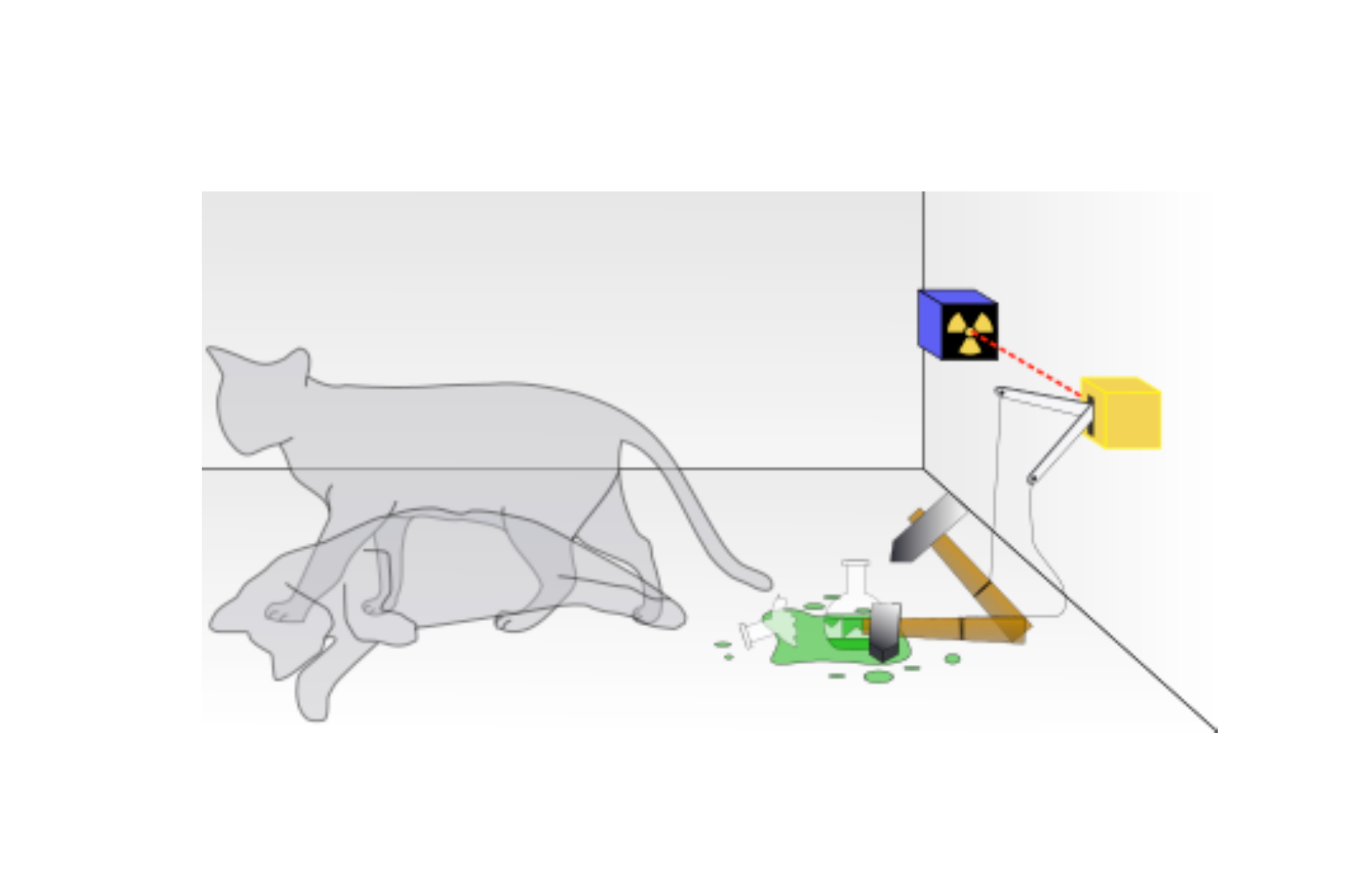} 
	      \vspace{-2\baselineskip}
	   \caption{Schrödinger's cat Gedankenexperiment}
	   \label{fig: Chat1}
	\end{figure}

If the probabilities indicate that a disintegration has a one in two chance of having taken place after one minute, Quantum Mechanics indicates that, as long as the observation is not made (or, more precisely, as long as there is no reduction of the wave packet), the atom is simultaneously in two states: intact and disintegrated. But the mechanism devised by Erwin Schrödinger links the state of the cat (dead or alive) to the state of the radioactive particles, so that the cat would be simultaneously in two states (the dead state \textit{and} the alive state), until that the opening of the box (that is to say, the observation) triggers the choice between the two states. In the meantime, it is impossible to tell if the cat is dead or not.	

Obviously, we can challenge this presentation of things, the passage from quantum formalism to natural language involving approximations.

Before the opening of the box, the cat is in a state which can be described, thanks to the bra-ket Dirac notation, as a «ket-vector» $|u\rangle$ of the complex Hilbert space of states $\mathcal {E}$.  As the kets form a vector-space, if $\lambda_{1}$ and $\lambda_{2}$ are some numbers representing the probability amplitudes, and $|u_{1}\rangle$ and $|u_{2}\rangle$ are two kets of $\mathcal{E}$, then :
\[
|u\rangle =\lambda _{1}\cdot |u_{1}\rangle +\lambda _{2}\cdot |u_{2}\rangle. 
\]
In the case of the cat, as the probability to find him dead or alive is 1/2, so  $\lambda _{1} = \lambda _{2} = √1/2 = 1/√2$, and we may represent its state as :
\[
|u_{cat}\rangle = \frac{1}{\sqrt{2}}.(|dead\rangle + |alive\rangle).
\]
So, to say that the cat is both dead and alive is a coarse simplification :  the probability amplitudes could be complex numbers and the shift from linear addition to linguistic conjunction is already an interpretation.

Moreover, we must recognize that this thought experiment, as it is described, is totally unrealizable. First, the technical conditions to preserve the superimposed state of the cat would be very difficult to realize. For the moment, they are, in fact, quite unrealizable for more than a few molecules.
Second, this thought experiment appears unrealistic even in principle. Indeed, we can never directly demonstrate, or measure, that the cat is both dead and alive, because the fact of knowing its state would necessarily cause the collapse of the wave function.

However, we can not get rid of this paradox so easily, especially if, instead of experimenting with a real cat, we start by testing smaller objects. Moreover, it is not entirely accurate that we can get no information about the cat and what's going on in the box without opening it.  For example, a mouse could then pass through this one and the probability of coming out could be measured. This weakened version of the experiment was actually performed at the atomic scale (by sending a photon to a superimposed atom in the excited and not excited state), and confirmed the predictions of the theory. Various other tests are in progress, where the cat is replaced by a beryllium ion trapped in a superimposed state, a superconducting quantum interference device where electons flow both ways around the loop at once when they are in the «cat's state», or again a piezoelectric «tuning fork» placed on a superposition of vibrating and non vibrating states, including about 10 trillion atoms. An experiment involving a flu virus has also been proposed and another involving a bacterium and an electromechanical oscillator is suggested. 

A variant of the Schrödinger cat's experiment is that of Wigner's friend. If the box is replaced by a fully enclosed laboratory in which a man  is locked up (this time a friend to whom the physicist Wigner, who remains outside, asks to do a Quantum Mechanics experiment), we meet again some problem : should we think that it is by opening the laboratory door, at the moment when Wigner himself becomes aware of the result of the measurement, that the reduction of the wave packet will take place and that one of the components of the state vector will be canceled? Should we not rather think that this will take place as soon as the observer inside the laboratory becomes aware of it?  But if there are two distinct observations, we shall have now two different state vectors – which is in contradiction with the assertions of Quantum Mechanics. Should we think, as Wigner seems to believe (\cite{Wig1}, that consciousness plays a special irreducible role in Quantum Mechanics? But how to explain the difference of appreciation of the two observers? Indeed, as long as Wigner did not open the laboratory door, the whole laboratory is, for him, in a state superposition, whereas, for his friend stayed inside and who has already taken knowledge of the result of the experiment, the wave function has already collapsed. Should we think there is a contradiction, or can we explain it, as Penrose claims (\cite{Pen}), by saying that the superposition of two conscious states is not paradoxical because, just as there is no interaction between the multiple quantum states of a particle, the superimposed consciousnesses here do not have to be conscious of each other?

All of this proves anyway that the interaction between quantum-level phenomena and more complex entities, closer to our macroscopic experience - which is not a fruit of our imagination - raises a lot of problems. So, the paradox remains, whatever one thinks of it,  and our starting question still remains too : do the quantum effects go to infinity or, as it would be reasonable to think, stop at the end of a certain moment or at a certain level? This is where the interpretations diverge.

In this context, if we do not content ourselves with the Copenhagen interpretation, initiated by Bohr and Heisenberg, and for which Quantum Physics does not describe an independent reality but the knowledge that we have of it, then different solutions may be presented. Going from the most extravagant to the most plausible, there would be:

\begin{enumerate}
\item The many worlds theory of Hugh Everett, which assumes that all complementary quantum states – going to infinity – are actually realized, meaning that the universe splits into multiple realities with every measurement.
\item The Wigner theory which assumes an influence of the consciousness of the physicist (or the one who opens the box). It would be the realization of this new state that would provoke, directly or indirectly, the collapse of the wave function. Relational and transactional interpretations, or speculations about the Zeno effect may be considered, in a certain way, as variants of Wigner's theory.
\item The theory of decoherence (defended by Murray Gell-Mann, Nobel Prize of physics 1963), which states that the superposition can be maintained only in the absence of interactions with the environment, which «trigger» the choice between the two states (dead or alive). In such a theory, the break is caused by physical interactions with the environment, so that the coherence is broken all the more quickly that there are more interactions, and thus practically instantaneously on the macroscopic scale where billions of billions of particles are involved. In other words, the superposition state can only be maintained for very small objects (a few particles).
\item Finally, theories seeking to modify Quantum Mechanics, either by adding a variable expressing initial conditions such as position (Bohm-de Broglie theory)(see \cite{Boh}), or by introducing additional hidden parameters in the quantum laws (e.g. gravitation for Roger Penrose).
\end {enumerate}

The many worlds  theory seems to ignore Occam's razor. Though not falsifiable, and undeniably elegant (since it stays closer to the Schrödinger equation - which it takes seriously - without introducing new artifacts), it is intellectually expensive and probably closer to science fiction than to real science. Wigner's hypothesis supposes that a simple look can have material consequences : as it happens, consciousness could kill, in the literal sense of the term. But what about animals? If we deny all consciousness, or at least any form of consciousness comparable to that of human beings, must we think that they themselves are capable of reducing the wave function? And otherwise, how are the currents flowing in the neurons of a human being different from those that excite the neurons of non-conscious animals to produce the collapse of the wave function? Wigner's anthropological solution encounters, in fact, numerous problems. Concerning, now, the theory of decoherence, one knows it is verified only in a small number of cases and by means of simplifying assumptions. Moreover, being essentially non-linear, it is difficult to deduce from the linear formalism of Quantum Mechanics. Finally, the decoherence being only the preliminary step to the measurement, not the measurement itself, there is a risk of only repelling the problem... to infinity. It seems, therefore, that there is room for further investigation : this is what Schrödinger's thought experiment has made it possible to highlight. And this is what physicists do everyday now, which however, as we shall see, surprise them in surprises, and put them in front of an extremely puzzling quantum reality. Let us introduce now some other paradoxes.

\section{Delayed choice experiments}

\subsection{Double-split experiment}

	As we know since Thomas Young (1801),  light can display characteristics of both classically defined waves and particles. In 1927, Davisson and Germer demonstrated that electrons show the same behavior, which was later extended to atoms and molecules (\cite{Eil}).
	
	In the quantum version of the double-slit Young's experiment, a coherent light source, such as a laser beam, illuminates a plate pierced by two parallel slits, and the light passing through the slits is observed on a screen behind the plate. The wave nature of light causes the light waves passing through the two slits to interfere, producing bright and dark bands on the screen, which seems to mean that light is only waves. 
	
	\begin{figure}[h] 
	   \centering
	      \vspace{-2\baselineskip}
	   \includegraphics[width=4in]{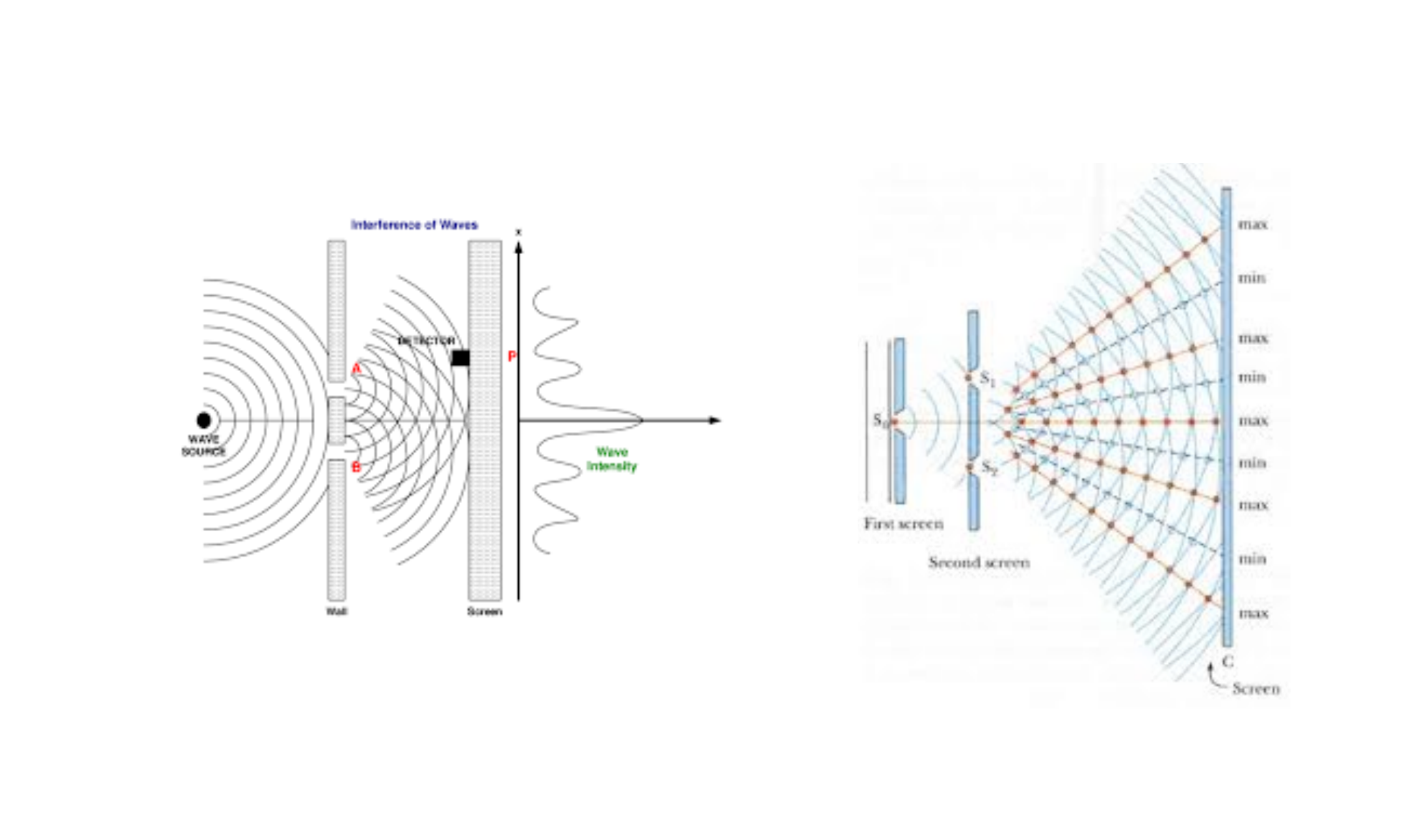} 
	      \vspace{-2\baselineskip}
	   \caption{The double-slit experiment}
	   \label{fig: slit1}
	\end{figure}
	
	However, the light is always found to be absorbed at the screen at discrete points, as individual particles, and the interference pattern only appears via the varying density of these particle hits on the screen. Furthermore, versions of the experiment that include detectors at the slits find that each detected photon passes through one slit (as would a classical particle), and not through both slits (as would a wave). So the conclusion was that particles do not form the interference pattern if one detects which slit they pass through (\cite{Asp3}). 
	
	For a long time, some scientists thought that the introduction of local hidden variables in Quantum Theory would allow them to precise which way photons or particles chose in the double-split experiment. But the refutation of the EPR argument in the 1980's forced them to give up that hope. 

\subsection{EPR and entanglement}

	In 1935, Einstein, Podosky and Rosen, anxious to highlight the incompleteness of Quantum Mechanics and the need to introduce hidden variables, proposed an experiment of thought that was actually done by the French physicist Alain Aspect in 1982. This experiment was based on the fact that, according to Quantum Mechanics, some particle systems have the special property of being «entangled», so that the knowledge of some observables characteristic of one of them lets you know the status of others instantly. This property was considered a defect for the authors, who wanted to show the incompleteness of Quantum Mechanics and its inability to describe an objective reality.
	
	More precisely, an entangled system is defined to be one whose quantum state cannot be factored as a product of states of its local constituents; that is to say, they are not individual particles but are an inseparable whole. In entanglement, one constituent cannot be fully described without considering the other(s). Note that the state of a composite system is always expressible as a sum, or superposition, of products of states of local constituents; it is entangled if this sum necessarily has more than one term.

Quantum systems can become entangled through various types of interactions. Entanglement is broken when the entangled particles decohere through interaction with the environment; for example, when a measurement is made.

An entanglement is produced when a subatomic particle decays into an entangled pair of other particles. The decay events obey the various conservation laws, and as a result, the measurement outcomes of one daughter particle must be highly correlated with the measurement outcomes of the other daughter particle (so that the total momenta, angular momenta, energy, and so forth remain roughly the same before and after this process). 

The special property of entanglement is maintained, even if we separate the said two entangled particles by a very long distance. For example, if we measure a particular characteristic of one of this pair (say, for example, spin), get a result, and then measure the other particle using the same criterion (spin along the same axis), we find that the result of the measurement of the second particle will match (in a complementary sense) the result of the measurement of the first particle, in that they will be opposite in their values.

A classical system would certainly display the same property. The difference is that a classical system has definite values for all the observables all along, while the quantum system does not. In a sense, the quantum system considered seems to acquire a probability distribution for the outcome of a measurement of the spin along any axis of the other particle upon measurement of the first particle. This probability distribution is in general different from what it would be without measurement of the first particle. This may certainly be perceived as surprising in the case of spatially separated entangled particles.

The distance and timing of the measurements can be chosen so as to make the interval between the two measurements spacelike, hence, any causal effect connecting the events would have to travel faster than light. But, according to the principles of special relativity, it is not possible for any information to travel between two such measuring events. It is not even possible to say which of the measurements came first. For two spacelike separated events $e_{1}$ and $e_{2}$, there are inertial frames in which $e_{1}$ is first and others in which  $e_{2}$ is first. Therefore, the correlation between the two measurements cannot be explained as one measurement determining the other: different observers would disagree about the role of cause and effect.

The conclusion of Einstein, Podolsky and Rosen was that Quantum Mechanics in an incomplete theory, and need hidden variables to explain the situation.

The EPR argument dates back to 1935 and its refutation, in the form of the rejection of local hidden variable quantum theories, dates back to 1982 and the famous Aspect experiments (see \cite{Asp1} for the project and \cite{Asp2} for the results), inspired by John Bell and Bernard d'Espagnat. 

Without going into details, let us say only that a very general statistical theorem, the Bell's theorem, to which all «realistic» theories satisfy, and in particular local hidden-variable theories, is systematically violated by Quantum Mechanics, which has been proven by experiments on both photons (polarized and unpolarized) and 2-spin particle systems. 

From the work of Greenberger, Horne, and Zelinger (\cite{GHZ}), it has been learned also that even more dramatic violations of local realism become possible for correlated particle systems. GHZ's initial argument focused on 3-particle systems, but $N$-particle generalizations are possible. And even if the experimental verifications have not yet been realized, it is very likely that the result will be the same as in the case of the Aspect experiments, and will once again give reason to Quantum Mechanics. 

Still other experiments, such as that of Hardy's thought experiment (\cite{Har},  \cite{Har1}), equally show that the predictions of Quantum Mechanics should be always verified and allow situations that local hidden variable theories prohibit. We take all these situations for granted.

\subsection{Wheeler's experiments}	

Indeed, in 1978, Wheeler (\cite {Whe1} had already devised a thought experiment to prove that the predictions of Quantum Mechanics were true and that the particles' behavior was not only independent of space, but also independent of time. Lacking adequate technology, this experiment could not be realized, but it was described as «the possibility to use the receptor at the end of the apparatus to record well-defined interference fringes»(\cite{Whe1}, which do not change anything to the result.

The distance of travel in a laboratory split-beam experiment might be about thirty meters and the time a tenth of a microsecond. But the distance could as well have been billions of light years and the time billions of years. 

In 1983, Wheeler proposes an astronomical extension of his experiment. Suppose two quasars separated by six seconds of arc and considered to be two distinct images of one quasar. Assume also that evidence has been found for an intervening galaxy G-I, roughly a quarter of the way from us to the quasar. 
	
	\begin{figure}[h] 
	   \centering
	      \vspace{-1\baselineskip}
	   \includegraphics[width=3in]{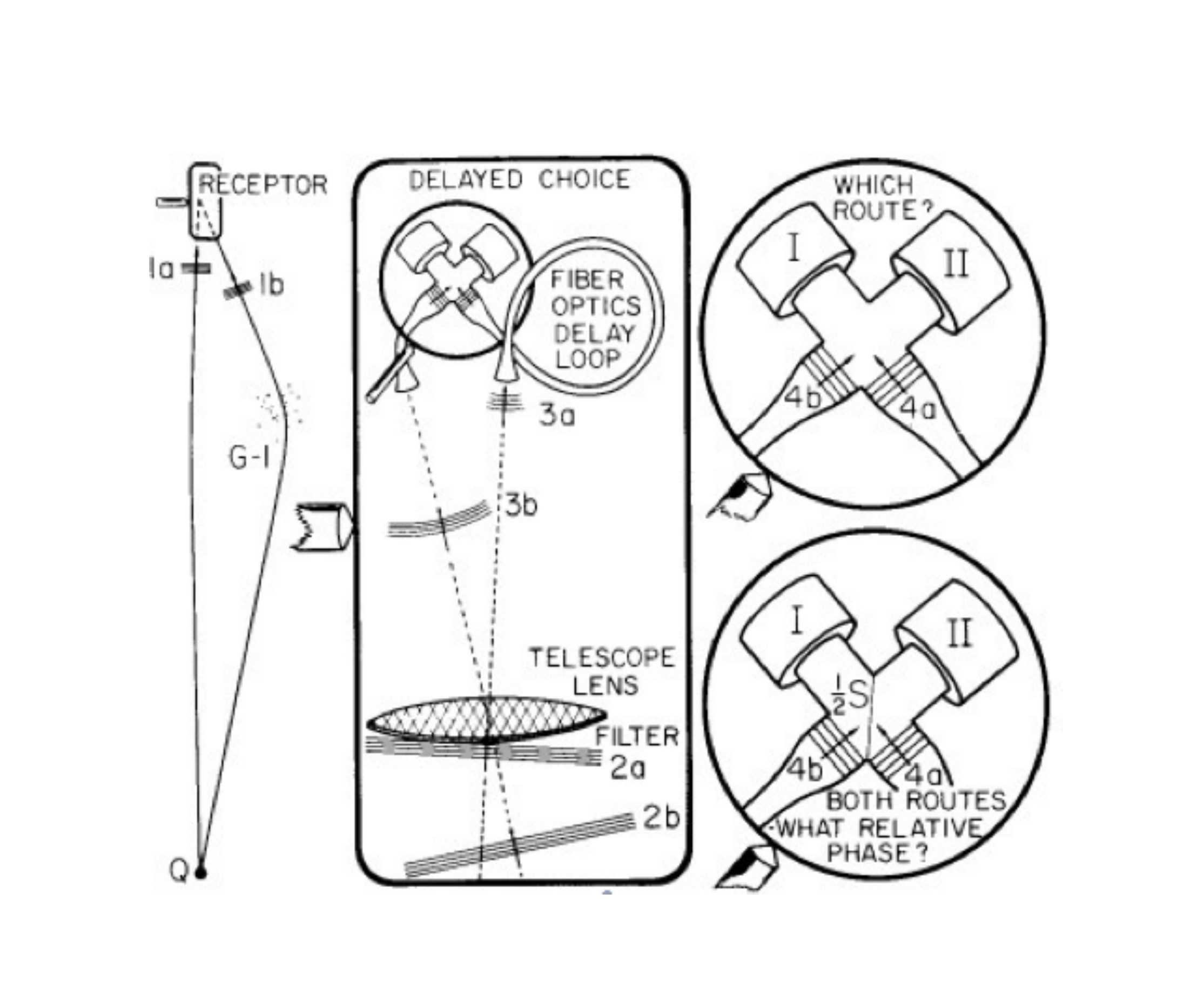} 
	      \vspace{-1\baselineskip}
	   \caption{Wheeler's cosmological experiment}
	   \label{fig: Whee1}
	\end{figure}

«Calculations indicate that a normal galaxy at such a distance has the power to take two light rays, spead apart by fifty thousand light years on their way out from the quasar, and bring them back together at the Earth. This circumstance, and evidence for a new case of gravitational lensing, make it reasonable to promote the spit-beam experiment in the delayed-choice version from the laboratory level to the cosmological scale» (\cite{Whe2}, 192).
	
	What are the consequences?

	At the end of the experiment, which takes time, light arrives on Earth and «we discover "by which route" it came with the one arrangement or by the other, what the relative phase is of the waves associated with the passage of the photon from source to receptor "by both routes" – perhaps 50,000 light years apart as they pass the lensing galax G-I. But the photon has already passed that galaxy billions of years before we made our decision.»(\cite{Whe2}, 192).

In 1988, Wheeler comes to the following conclusion : «There is an inescapable sense in which we, in the here and now, by a delayed setting of our analyzer of polarization to one or another angle, have an inescapable, an irretrievable, an unavoidable influence on what we have the right to say about what we call the past» (\cite{Whe3}).

Suppose now that the photon comes from even further, say, from the primordial universe – in any case, as soon as the photons are released –, it means that close decisions reach backward in time in their consequences, back even to the earliest days of the universe.

Of  course - Wheeler recognized - for the Copenhagen interpretation, it is a wrong way of speaking. According to Bohr, «no elementary phenomenon is a phenomenon until it is a registered (observed) phenomenon». It means that phenomena, called into being by decisions, reach into the present from  billion of years in the past.

Since 1982, a number of concrete so-called «delayed-choice» experiments have tended to confirm this result, which is now admitted by leading physicists and mathematicians, although they do require careful analysis. For some of them, perhaps it is better to speak, as Wheeler also does, of «an elementary act of creation». As he writes, «it is wrong to think of that past as "already existing in details". The "past" is theory. The past has no existence except as it is recorded in the present.»(\cite{Whe3}, 194).

But though puzzling the situation is, it is a fact : «By deciding what questions our quantum registering equipement shall put in the present, we have an undeniable choice in what we have the right to say about the past»(\cite{Whe3}, 194).

In 2006, the results of one of the first realizations of Wheeler's experiment by Aspect and the team of ENS Cachan in France, were again in agreement with the predictions of Quantum Mechanics. We just report here the conclusion of the article :

«Our realization of Wheeler’s delayed-choice GedankenExperiment demonstrates beyond any doubt that the behavior of the photon in the interferometer depends on the choice of the observable which is measured, even when that choice is made at a position and a time such that it is separated from the entrance of the photon in the interferometer by a space-like interval. In Wheeler’s words, since no signal traveling at a velocity less than that of light can connect these two events, “we have a strange inversion of the normal order of time. We, now, by moving the mirror in or out have an unavoidable effect on what we have a right to say about the already past history of that photon"» (see \cite{Jac}, 2-3). 

As the authors said, with a bit of understatement, this surprising situation seems to reveals «some tension» with Relativity. 

This cautious conclusion has however drawn the wrath of a recent interpreter, who considers that this is again a wrong way of speaking. Remaining as close as possible to the Copenhagen interpretation, Hervé Zwirn (\cite{Zwi}) tries to solve the problem by considering that, since we can not speak of phenomena (and thus of trajectories or positions, waves or particles) before they are duly recorded, each observation, linked to a conscious observer, corresponds in sum to a way of seeing the world. As no observation is contradictory with another, this solution is named by him «friendly solipsism». 

No need to be a great cleric to see that this interpretation is very much like Hugh Everett's theory of the many worlds, except that the worlds in question do not exist outside of the consciousnesses of individuals. This looks like a kind of revival of  the old leibnizian monadology.

But Wheeler had bothered to eliminate such solutions. «Consciousness has nothing whatsoever to do with the quantum process. We are dealing with an event that may itself known by an irreverslble act of amplification, by an indelible record, an act of registration. Does that record subsequently enter into the "consciousness" of some person, some animal or some computer? (...) Then that is a separate part of the story, important but not to be confused with "quantum phenomenon".»(\cite{Whe2}, 196).		

\subsection{Scully's experiment}

Another delayed-choice experiment, involving entanglement has been made since the publication of Wheeler's articles. One among the most spectacular, probably, has been carried out by the American physicist Marian Scully (\cite {Scu}), specialized in quantum optics. It belongs to a category of experiments named «delayed-choice quantum eraser experiments», first performed by Yoon-Ho Kim, R. Yu, S. P. Kulik, Y. H. Shih and Marlan O. Scully, and reported in early 1999 (\cite{Kim}, see also \cite{Scu}). This experiment was designed to investigate peculiar consequences of the well-known double-slit experiment in Quantum Mechanics, as well as the consequences of quantum entanglement.
	
	In the double-slit experiment, as we know, if particle detectors are positioned at the slits, showing through which slit the photon goes, the interference pattern disappears. However, in 1982, Scully and Drühl proposed a «quantum eraser» to obtain which-path information without scattering the particles or otherwise disturbating them. Rather than attempting to observe which photon was entering each slit, they proposed to «mark» them with information that, in principle, would allow the photons to be distinguished after passing through the slits. Of course, the interference pattern does disappear when the photons are so marked. However, the interference pattern will reappear if the which-way information is further manipulated after the marked photons have passed through the double slits to obscure the which-way markings. Let us see in details how does it work.	

	\begin{figure}[h] 
	   \centering
	      \vspace{-2\baselineskip}
	   \includegraphics[width=4in]{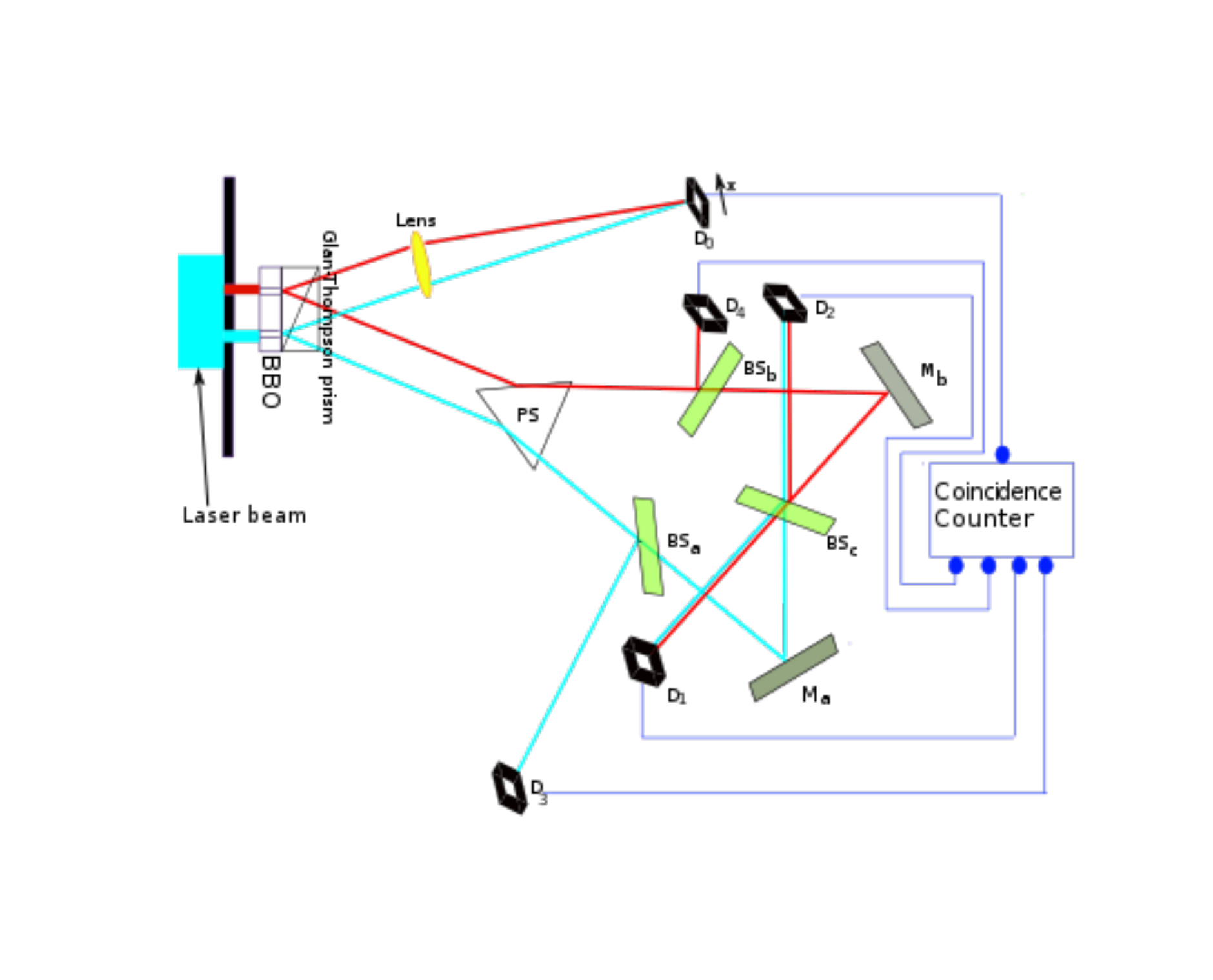} 
	      \vspace{-2\baselineskip}
	   \caption{Scully's experiment}
	   \label{fig: Scully1}
	\end{figure}

	The experiment begins as a conventional two-slit experiment. However, after the slits, spontaneous parametric down-conversion (SPDC) is used to prepare an entangled two-photon state. This is done by a crystal BBO (beta barium borate) that converts the photon (from either slit) into two identical, orthogonally polarized entangled photons with 1/2 the frequency of the original photon. The paths followed by these orthogonally polarized photons are caused to diverge by a prism.
	
	One of the photons, referred to as the «signal photon»  continues to the target detector called $D_{0}$. A plot of «signal photon» counts detected by $D_{0}$  can be examined to discover whether the cumulative signal forms an interference pattern.
	The other entangled photon, referred to as the «idler photon», is deflected by prism PS that sends it along divergent paths depending on whether it came from slit A or slit B.
	
	Somewhat beyond the path split, the idler photons encounter beam splitters $BS_{a},\ BS_{b}$, and $BS_{c}$ that each have a 50\% chance of allowing the idler photon to pass through and a 50\% chance of causing it to be reflected. $M_{a}$ and $M_{b}$ are mirrors.

	The beam splitters and mirrors direct the idler photons towards detectors labeled $D_{1}, D_{2}, D_{3}$ and $D_{4}$. We can see that:
	
	\begin{enumerate}
	\item If an idler photon is recorded at detector $D_{3}$, it can only have come from slit B.
\item If an idler photon is recorded at detector $D_{4}$, it can only have come from slit A.
\item If an idler photon is detected at detector $D_{1}$ or $D_{2}$, it might have come from slit A or slit B.
	\end{enumerate}
	
	The optical path length measured from slit to $D_{1}, D_{2}, D_{3}$, and $D_{4}$ is 2.5m longer than the optical path length from slit to $D_{0}$. This means that any information that one can learn from an idler photon must be approximately 8ns later than what one can learn from its entangled signal photon.

	Detection of the idler photon by $D_{3}$ or $D_{4}$ provides delayed «which-way information» indicating whether the signal photon with which it is entangled had gone through slit A or B. On the other hand, detection of the idler photon by $D_{1}$ or $D_{2}$ provides a delayed indication that such information is not available for its entangled signal photon. Insofar as which-way information had earlier potentially been available from the idler photon, it is said that the information has been subjected to a «delayed erasure».
	
	The results of the experiment are the following ones:
	
	\begin{enumerate}
	\item When the experimenters looked at the signal photons whose entangled idlers were detected at $D_{1}$ or $D_{2}$, they detected interference patterns.
	\item However, when they looked at the signal photons whose entangled idlers were detected at $D_{3}$ or$D_{4}$, they detected simple diffraction patterns with no interference.
\end{enumerate}

	The significance of these results is similar to that of a classical double-slit experiment, since interference is observed when it is not known from which slit the photon originates, while no interference is observed when the path is known.

	However, what makes this experiment possibly astonishing is that, unlike in the classic double-slit experiment, the choice of whether to preserve or erase the which-path information of the idler was not made until 8ns {\it after} the position of the signal photon had already been measured by $D_{0}$.
	
	So, everything goes as if the idler photon have «said» to its entangled signal photon, just before it reaches the detector $D_{0}$ :
	
	– In the first case, that the observer knows which way it followed and so, that if has to behave as a particle and let just a hit on the screen;
	
	– In the second case, that the «which way» information is lost and that it has now to behave as a wave and interfere with itself.
	
	Scully merely reports the results of the experiment without interpreting them, but it is clear that it fully checks the predictions of Quantum Mechanics as well as Wheeler's assertions.	
	
\section{Performing the cosmological experiment}

In the previous experiment, and all those of the same type performed in a laboratory, the distances, as Wheeler already observed, were obviously very short and the delay between the two observations no longer than a few nanoseconds. But, as we have seen, Wheeler proposed a cosmological experiment  which remained also a GedankenExperiment until 2017. 

At that date, a new experiment, more in line with what Wheeler had imagined, came into play.

The physicist Paolo Villoresi and his colleagues at the University of Padua (see \cite{Vil1} and \cite{Vil2}) sought to know what Wheeler's experience would be if it was done with a laser beam over a long distance. As the researchers explained in an article published in {\it Science Advances}, they used the instruments available at the Italian Matera Laser Ranging Observatory (MLRO) to create a giant Mach-Zender interferometer by connecting the observatory and satellites in low Earth orbit a laser beam. Despite a photon distance of about 3,500 km, Wheeler's delayed choice experiment gave the same results as its much smaller terrestrial counterparts. Of course, this reinforces our confidence in the universal validity of Quantum Physics equations, but asks also a puzzling question : Have we, yes  or no, acted on the past?

\begin{figure}[h] 
	   \centering
	      \vspace{-2\baselineskip}
	   \includegraphics[width=4in]{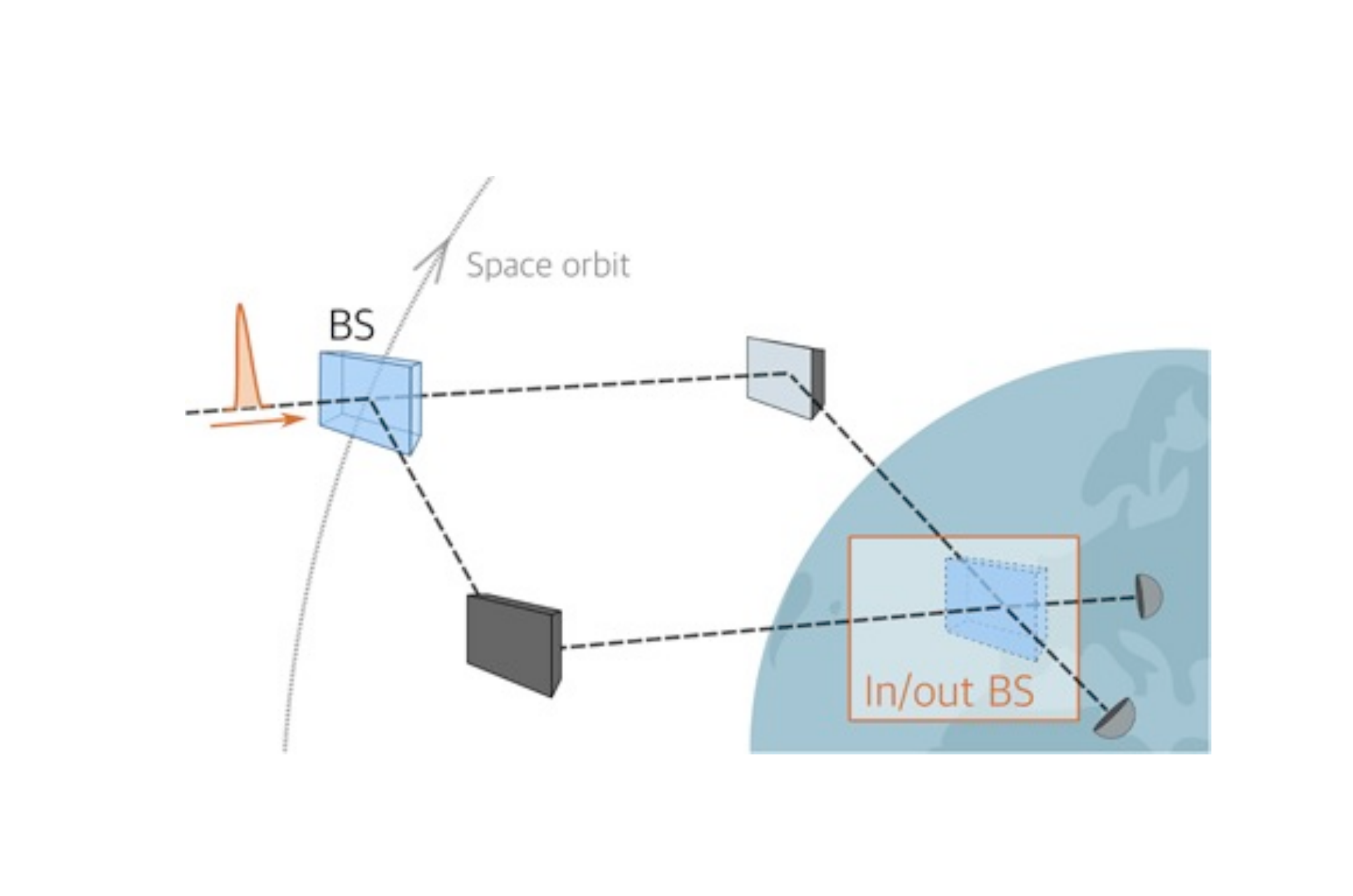} 
	      \vspace{-2\baselineskip}
	   \caption{Villoresi's cosmological experiment}
	   \label{fig: VIlloresi11}
	\end{figure}

We do not want to comment this experiment in detail. We merely describe the previous figure. A photon wave packet enters the first BS of an interferometer, which extends along thousands of kilometers in space. The interferometer can be randomly arranged according to two configurations that correspond to the presence or absence of the second BS (in/out BS) located on Earth. Following Wheeler’s idea, the configuration choice is performed when the photon has already entered the interferometer. In the actual implementation, the interferometer begins and terminates on the ground, extending up to the target satellite, and the measurement choice performed on ground is space-like separated from the photon reflection by the satellite. The results are of the same kind as in the Scully's experiment.

\section{Retrocausality?}

In the standard view conveyed by the Copenhagen interpretation, if a photon in flight is interpreted as being in the so-called «superposition of states», i.e. if it is interpreted as something that has the potentiality to manifest as a particle or wave, but during its time in flight is neither, then there is, in principle, no time paradox. Far more, the common element in the previous experiments is that it is the link between some separation apparatus and an appropriately spatially-positioned detector (see \cite{Fey}) that induces the collapse of the state vector to an eigenstate, so that the detector will only register one of the eigenstates. The motto is in short: «go around, there is nothing to see».

In fact, I am afraid things be more complicated and I would want to propose some new «GedankenExperimenten», even if they are not really plausible and, as a result, will never be likely performed.

I will start with some observations about the past and the so-called «historical» facts.

From the Trojan War to the Gulf War, military history has been marked as much by luck and error as by heroism and genius. Sometimes very small events have even been decisive in the outcome of a battle. That is what Erik Durschmied (\cite{Dur},  in a book I reported elsewhere (\cite{Par}),  called «the logic of the grain of sand» :  a fact or behavior, totally unpredictable, would sometimes upset the objective data of a conflict and allow the triumph of one of the two camps over the other.

Here I would like to go down a few orders of magnitude and consider what could happen to some past events if, as in the case of the Schrödinger's cat, it turned out that we could - if only in thought - make them depend on microphysical realities : no longer human actions or even «grains of sand» in this case, but a few photons or wave packets whose behavior, after all, may have been as well important at some time : somebody may be dazzled by a beam of light and, because of that, could have reacted inappropriately and made a bad decision.

Of course, no chance to recover photons from Waterloo or Pearl Harbour and to modify the outcomes of these battles. But the experiment remains plausible if applied to a near past and to very simple stimuli. It is known that, in the best visibility conditions, about 400 photons hit the human retina. However, as the threshold duration of the excitation is only a fraction of a second, a few photons are enough to trigger the phenomenon of visual sensation, as shown by experiments conducted by Hecht, Shaer and Pirenne in 1941, then by Van der Velden in 1944. The latter set the threshold, that is to say the intensity which causes 50 \% positive responses to 2 photons corresponding to the wavelength of 507 m$\mu$, provided that they arrive on the optical ganglion with between them an interval of time lower than a certain $\tau$, equal to the rate of relaxation of the ganglion neuron (see \cite{Gou}, 112-114). It would therefore be sufficient that, by a suitable device, these 2 photons could be entangled and recovered in appropriate detectors to make us able to act on the past of a human perception and, as it may happen, to modify, as was said above, the evolution of a past event.

Obviously, such a pseudo-experiment is pure speculation, and it is doubtful that it can ever be realized. What I want to say is that we can not be content with what the Copenhagen interpretation tells us, because the vagueness in which it leaves the behavior of the particles between the moment they are emitted by the source and the one where they are detected - the famous superposition of states - no longer holds when precisely linking this behavior to macroscopic realities.

Now it is a fact that microphysical entities, whatever they may be, not only have quantum effects but, step by step, effects at higher levels. For the purpose of the demonstration, I chose historical facts, but I might as well have chosen - and probably with greater chances of being plausible - biological realities. Who knows if the behavior of elementary particles does not have effects on certain genetic mutations? Everything suggests that this is probably the case, and one can then imagine other thought experiments where present manipulations could command significant bifurcations in the near past evolution of the living. I am not saying that to admit the existence of a retrocausality is to think that we could remotely modify the length of the nose of Cleopatra, for God knows what consequences on the behavior of Caesar or Pompey, the continuation of the Roman history and, more generally, the rest of world history in general. But, here again, if we could isolate some genetic effects of entangled particles, experiments on generations of viruses or bacteria could be considered.

All that only proves something is wrong in our present Quantum Physics. Despite the stoic philosopher Cleanthes ($\sim$-330 -232) who seems to have been one of the few to deny the irrevocability of the past, we can not  admit such an instability and no more accept an infinite multiplication of parallel (orthogonal, in fact) universes à la Hugh Everett. There is a very little chance this might be true. 
But, on the contrary, I do not think that we should take literally Niels Bohr's famous phrase simply identifying phenomena with registered (or observed) phenomena. It is simply not tenable and the existence of proven historical realities whose existence remains partly hidden and of which we only become acquainted long after they have occurred, in the form of direct or indirect traces, when it is not fossils, is sufficient to show the adventurous nature of such a statement. 

Scientists who follow a little too closely Niels Bohr's idealistic assertions about the existence of phenomena would do well to think twice. Again to support our argument and to make clear what we mean, let us put the problem in the field of medicine. 

Everyone will agree, I think, that humanity is subject to very real pathologies since it exists, but, of course, these are only discovered, named and understood in the course of history. For example, whooping cough, cholera, scarlet fever or AIDS existed long before they were discovered, and in fact long before we understood they were just examples of, respectively, Gram-negative cocci, Gram-negative bacilli, Gram-positive cocci and viral diseases. On the other hand, the deviations of the modern pharmaceutical industry lead sometimes  to the creation of «ad hoc» diseases, the reality of which is not proven. However, particularly important populations are generally targeted to maximize potential revenue : these are examples of «disease mongerings» (see, for example, alopecia, restless leg syndrome, testosterone deficiency, hypoactive sexual desire disorder, etc.). But who would dare to argue that these pseudo-diseases have the same reality as the previous ones? The former are real diseases, validated by biochemistry; the later are only socio-linguistic constructions. Therefore, it is not enough to name a phenomenon so that it corresponds to a real existence. In the field we talk about, the difference between the two kinds of diseases is precisely that of good and bad medicine. 

Indeed, there is no good and bad physics. Quantum Mechanics has just to be as good physics as the theory of relativity. Consequently, the least we can recognize is that it must not create social or mental chimeras. It has to describe objective phenomena that must be objectively reported.

So, where is the solution? Must we think that we live in a «participatory universe» (see \cite{Zei1}, \cite{Zei2}) or all this is only misinterpretations?

\section{Discussion}

Delayed-choice experiments seem to suggest not only that the future may influenced the past but that, symmetrically, the past will allow us to know about the future before it happens. 

As the 2 particles $P_{1}$ and $P_{2}$, actually correlated, are separated in space-time by a spacelike interval, we could therefore deduce that, since they could not influence each other by means of a signal in the time interval considered, communication - if there is communication - should have followed another path in the Einsteinian space-time, which is a fully deployed block.

\begin{figure}[h] 
	   \centering
	      \vspace{-2\baselineskip}
	   \includegraphics[width=4in]{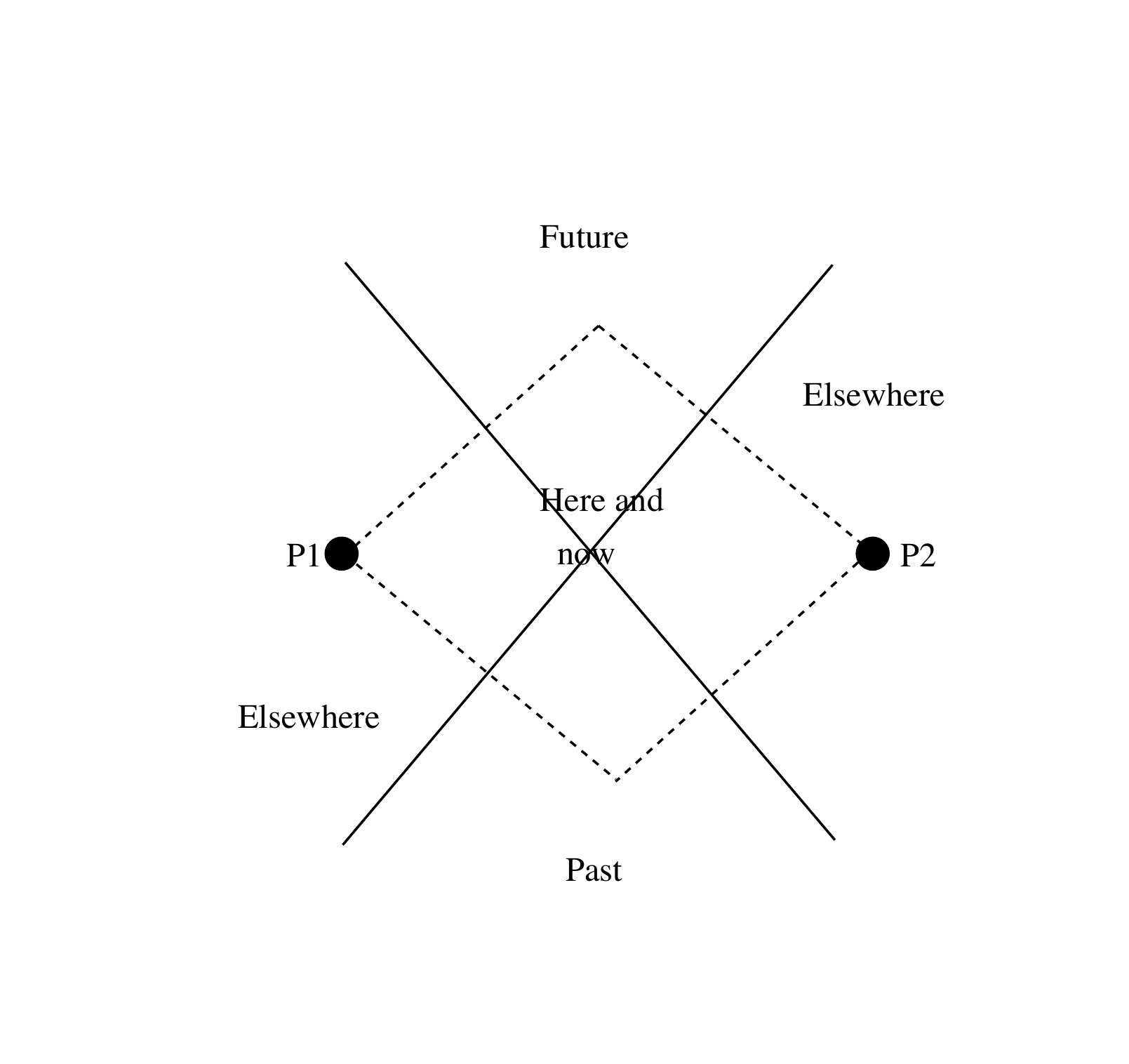} 
	      \vspace{-2\baselineskip}
	   \caption{Einsteinian space-time and quantum correlations}
	   \label{fig: Costa1}
	\end{figure}

Let us recall, however, that the velocity of a signal, according to the theory of relativity, is bounded superiorly by the speed of light, and that the particles have been precisely chosen in such a way that no signal of this kind can exist, in any case, in the time interval considered by the experiment. And as the particles could not have influenced each other through the present, it therefore remains to be assumed that they have had some communication «by taking a relay either in the past or in the future» (\cite{Cos4}, 68). We shall see that some physicists like Costa de Beauregard have supported this thesis in these words (see, in particular, \cite{Cos3}, 429-431), bringing this characteristic zig-zag closer to the Feynman diagrams assimilating a positron to an electron going back in time. On the contrary, other physicists and, in general, epistemologists, consider this interpretation as completely abusive and appropriate to reinforce the most irrationalist theses (see \cite{Gra}, 238-239).

In fact, the experiments reported above, reveal a total reversibility. The interference pattern can only be seen retroactively once the idler photons have been detected and the experimenter has had information about them available, when some signal photons were matched with idlers gone to particular detectors.

So, for the mainstream interpretation, there is no retrocausality at all but only correlations of entangled photons' observables that can be described in two ways : in the historic order and we say that the position at $D_{0}$  of the detected signal photon determines the probabilities for the idler photon to hit either of $D{1},\ D{2},\ D{3}$ or $D{4}$; on in the reverse order, and we say that the detector $D_{i}$, at which the idler photon is detected, determines the probability distribution at $D_{0}$ for the signal photon.

But, symmetrically, no advanced causality, and so, no knowledge of the future is possible : as the total pattern of signal photons at the primary detector never shows interference, so it is not possible to deduce what will happen to the idler photons by observing the signal photons alone. Indeed, the delayed-choice quantum eraser can not communicate information in a retro-causal manner because it would necessitate a process that go faster than the speed of light, which is totally excluded.

Another argument in favor of this interpretation is a theorem proved by Philippe Eberhard, which shows that if the accepted equations of relativistic quantum field theory are correct, it should never be possible to experimentally violate causality using quantum effects (see \cite{Ebe}, \cite{Ell}). So, for most authors, delayed-choice experiments reveal, once more, the well-know effects of non-locality and that is all. We agree that those experiments do not prove we can know anything about the future. 

Indeed, for relativity theory, as we have recalled, the future is already over there, in some part of space-time. It is probably the reason why, despite Eberhard's proof, some physicists have speculated that these experiments might be changed in a way that would be consistent with previous observations, yet which could allow for experimental causality violations (see \cite{Cram1}, \cite{Cram2}, \cite{Wer}).

I am not suggesting that these interpretations are necessarily the right ones. But I affirm that we can no longer be satisfied today with explanations that are just dogmas imposed by a purely pragmatist, instrumentalist and, ultimately, positivist philosophy which, when in difficulty with experimentation, tries to escape in the pleasant manner of Wheeler, by making the particle «a great smoky dragon that is only sharp where it enters the interferometer and where it leaves the interferometer biting the detector». Metaphors have nothing to do in physics and the collapse of the wave function would have also to be explained.

Of course, for anyone who knows a little physics, «great smoky dragon» means «non-locality», and «non-locality» refers to the formalism of Quantum Mechanics and a strict reading of Hermitian reversibility. But the word only points in the direction of a problem: what is the meaning of this non-locality? A new field hitherto unnoticed? But where are its characteristics and how to experiment with it? We should get variables or parameters that can be measured. But mainstream Quantum Mechanics, supposedly complete, forbids them and, in fact, we must recognize that they can not exist in the form of hidden local variables. Anyway, the non-locality is only a word, and the «veiled reality» to which some physicists like Bernard d'Espagnat appeal to explain it, recalls, without going much further, those «general principles of the corporal nature and mechanics» that Leibniz in his {\it Discourse of metaphysics} (§18)  called «metaphysical», and which, according to him, were «forms or indivisible natures» that cause the appearance (see \cite{Esp}, 271 and \cite {Par1}, 65).

Bernard d'Espagnat himself, although he always rejected the idea that his «veiled reality» could replace religious beliefs, was not far from seeing God behind this one.  The deep french philosopher Raymond Ruyer had however, in his time, rejected such a thought: "It is naive", he wrote, "to imagine that beyond space-time and observables, begins \textit {immediately} the reign of God, of the One, of the Eternal, and the competence of the theologian. The trans-spatial is not the supernatural" (\cite{Ruy1}, 86). Behind the world of appearances exists that of quantum physics and this vast domain of trans-space. But "one does not find in it the ubiquity, the eternity and the divine unity, but a sort of relative ubiquity, of relative eternity, of relative unity" which is in fact that of consciences and memories without which these phenomena would lose all meaning. Without doubt, just as rationalism, for Bachelard, became, with the scientific revolutions of the twentieth century, a surrationalism, so could the expanded naturalism of contemporary science be called, if it did not lead to confusion, "trans-naturalism". In this sense, the unobservable trans-spatial could gradually merge into an unknowable trans-natural. But, as far as we can speak of it, the unknowable divine would be at best a limit.

\section{Free-Will theorem and other connected propositions}

Assume now, for a while, that a kind of connection between the future and the past may exist : we do not speak of «retrocausality» and do no more suppose that the past might be a posteriori modified, except perhaps at a level with very limited macroscopic consequences, i.e. in the limit of indetermination allowed by the Heisenberg principle. What consequence on the alleged human freedom? Does it mean we are free or not? 

For different reasons, the various attempts to apply modern physics to living beings (\cite{Fan}, \cite{Jor}) had not been very convincing. As Raymond Ruyer once remarked, «contemporary physicists who, taking micro-physical indeterminism seriously, have spoken of the "freedom of the electron" by relating it to human freedom, have not had very good press»(\cite{Ruy}, 153). Remaining, on the contrary, as close as possible to the teachings of quantum physics, this philosopher, aware of the advances of this science, had introduced, for every entity existing in the manner of a «subject», the notions of «absolute flying-over domain», of «trans-spatial region» and of «finality», which he regarded as intrinsic characteristics of all physical beings.

If the first two characteristics seemed to correspond to the phenomena of superposition of states and entanglement, the idea of finality seemed to overstep the possibilities of a particle. But Ruyer supported the idea that freedom does not consist in producing movements without reason. And if it seemed difficult, in appearance, to find in microphysics, the equivalent of the «end» characteristic of the free activity, which always seeks to reach a final state, optimal, according to a norm, we could notice that reduced to the essentials, this character only returned to this: an activity whose changes could not be related to causes {\it a tergo}, but were defined by a final state in the most general sense of the word. Thus, Ruyer observed, «an atom of hydrogen is self-made incessantly. It can no more "be there" once and for all, than a living being or a social institution. Since it is nevertheless possible to characterize it as a hydrogen atom, it must therefore obey a norm, and its nature must be a {\it physis} in the etymological sense of the word»(\cite{Ruy}, 157) .

Curiously, some recent physicists have re-investigated the problem.

The «free will theorem» of John H. Conway and Simon B. Kochen states that if we have a free will in the sense that our choices are not a function of the past, then not subjected to certain assumptions, so must some elementary particles. Conway and Kochen's article was published in {\it Foundations of Physics} in 2006 (see \cite{Con1}). In 2009 they published a stronger version of the theorem in the {\it Notices of the AMS} (see \cite{Con2}). And later, Kochen elaborated some details (\cite{Koc}).

The proof of the theorem as originally formulated relies on three axioms, which Conway and Kochen call «fin», «spin», and «twin», the last two being able to be verified experimentally.

$Fin$: There is a maximal speed for propagation of information (not necessarily the speed of light). Such an assumption rests upon causality.

$Spin$: The squared spin component of certain elementary particles of spin one, taken in three orthogonal directions, is a permutation of (1,1,0).

$Twin$: It is possible to «entangle» two elementary particles and separate them by a significant distance, so that they have the same squared spin results if measured in parallel directions. This is a consequence of quantum entanglement (full entanglement is not necessary for the twin axiom to hold : it is sufficient but not necessary).

In their later paper, «The Strong Free Will Theorem», Conway and Kochen replace the $Fin$-axiom by a weaker one called $Min$, thereby strengthening the theorem. $Min$ asserts only that two experimenters separated in a space-like way can make choices of measurements independently of each other. In particular it is not postulated that the speed of transfer of all information is subject to a maximum limit, but only of the particular information about choices of measurements. More recently, Kochen argued that $Min$ could be replaced by $Lin$ - experimentally testable Lorentz Covariance. 
The free will theorem states:

Given the axioms, if the two experimenters in question are free to make choices about what measurements to take, then the results of the measurements cannot be determined by anything previous to the experiments : the outcome is open.

If the outcome of an experiment was open, then one or two of the experimenters might have acted under free will.

Since the theorem applies to any arbitrary physical theory consistent with the axioms, it would not even be possible to place the information into the universe's past in an {\it ad hoc} way. The argument proceeds from the Kochen–Specker theorem, which shows that the result of any individual measurement of spin was not fixed independently of the choice of measurements. As stated by Cator and Landsman regarding hidden-variable theories, «there has been a similar tension between the idea that the hidden variables (in the pertinent causal past) should on the one hand include all ontological information relevant to the experiment, but on the other hand should leave the experimenters free to choose any settings they like»(see \cite{Cat}).

The conclusion to be drawn from the theorem seems to be  that «science does not support determinism».  Quantum Mechanics obviously shows that particles do indeed behave in a way that is not a function of the past. But some critics have argued that the theorem applies only to deterministic models when others have claimed that the indeterminism Conway and Kochen think to have established was already assumed in the premises of their proof. 

Thus, no decisive argument exists in favor of an elementary «freedom» of particles, nor of any relation between it and our supposed human «freedom», moreover envisaged only here in the form of a cartesian «free will». Maybe we are free to make (or not) physical experiments, but this does not mean that nature is free to do anything.  Ruyer's definition of freedom as a finalist activity was surely a better one, but the difference in level between particle physics and the sciences capable of explaining the behavior of the living (and especially of the human living) is such that it renders all comparisons very delicate..

\section{Acommodating retrocausality with Free will?}

Indeed, there is effectively a tension between freedom and causality, especially when it appears, no matter what, in the form of retrocausality. This could be, of course, a false problem, especially if we reject the possibility of retrocausal actions. Nevertheless, a number of physicists ask the question. We shall try, here, to give an account of their arguments.

Retrocausality is not, in fact, a new idea. Before Wheeler'experiments and Aharonov two-vector approach to Quantum Mechanics (voir \cite{Aha}), i.e. as early as 1947, and explicitly in 1953, Olivier Costa de Beauregard (\cite {Cos1}) - as said above - had already evoked such a hypothesis to which he had then held his whole life. However, his presentation has been subsequently modified in the course of time. In one of his latter books (\cite{Cos2}), the EPR situation was read in terms of conditional probabilities, so that predictive calculations when one follows the direct order of events, and retroductive calculations, when one follows the inverse order, are formally identical. Consequently, the notion of causality is simply reduced to the notion of law, Hermitian reversibility plays a full role, and we are no longer very far from the Copenhagen interpretation. The explanation of this reversibility is just connected to the CPT-invariance.

The ideas of Costa de Beauregard were founded on Loschmidt observation that Boltzmann could not deduce his H-theorem asserting that entropy is always increasing in a closed system from the laws of dynamics which are completely symmetric relatively to the variable $t$ (for some solutions, see \cite{Aha1}, \cite{Sta}).  A more puzzling problem comes with some models of non-standard cosmology like the Gold universe (see \cite{Gol1}, \cite{Gol2}, \cite{Gol3}; and also \cite{Pri}, \cite{Mer}). In this kind of model, the universe expands for some time, with increasing entropy and a thermodynamic arrow of time pointing in the direction of the expansion but, after it reaches a low-density state, it recontracts. Then entropy decreases, pointing the thermodynamic arrow of time in the opposite direction, until the universe ends in a low-entropy, high-density Big Crunch. This suggests the universe will become more orderly after the moment of contraction. Of course, the Gold model is necessary linked to the possibility of retrocausal change, and so, questions arise concerning the preservation of information in states of decreasing entropy, where a reverse causation should exist.

But it is one thing to say that time does not exist in microphysics where, apart from the case of mesons, there is nowhere irreversibility, and it is another thing to consider squarely a total reversal of the arrow of time and causal relationships. One could estimate in the 1960s that cosmological retrocausality seems to be independent of quantic retrocausality. But it is no longer the case in 2019 where cosmology became quantic cosmology.

Retrocausality may be accommodated with free will insofar we stay in the frame of Quantum Physics stricto sensu. But what happens, for example, in a Gold universe? Must we think that, after the low density point, the course of history is totally reversed as in «The Curious Case of Benjamin Button», this fantastic novel of Francis Scott Fitzgerald (\cite{Fit}), whose David Fincher made a movie? One will say that, as everybody and everything will be like the heroe, nothing is change. But what happens to men (and things) born or made in the world just before the cusp and submitted, first to an increasing entropy, and then, to a decreasing one? Can we save a free will in such a universe? Indeed, it would be very strange to have intentions and to make projects in a universe where, instead of being performed, they suddenly collapse, especially since all the things on which we could count or on which our intentions or projects were based, gradually disappear from our sight... A solution would be to substitute to the Gold universe a multiverse where decreasing entropy concern a world which is not ours. That already was the solution of Boltzmann to accommodate entropy with reversibility of dynamical laws.

Proponents of David Bohm's theory (see, for instance, \cite{Wha}) do not take into account such situations, preferring models which, including apparent particle-like behavior and allowing controllable constraints on unknown past fields, are retrocausal but not retro-signaling and respect the conventional block universe viewpoint of classical spacetime.

\section{Conclusions}

Quantum physics obviously reveals a quite surprising world with very strange situations, as those like superposition of states, EPR argument and delayed-choice experiments, all situations which confine to philosophy, especially to the philosophy of time, but not only : the collapse of the wave function and the paradox of measurement, the existence of non-locality, the possibility a subtle form of retrocausality seem to point to an entity or a structure which knows neither time nor space. Strangeness is still increasing with quantum cosmology, which is concerned with the beginning (resp. the end) of the universe, brings to its peak the tension between time and the absence of time and ask questions about human freedom.

Some take the plunge and do not hesitate to venture on the paths of metaphysics, even of theology (see \cite{Crai1}) - sometimes founded on scientific confirmations : for example, the Borde-Guth-Vilenkin Theorem (\cite{Bor}), a cosmological theorem which deduces that any universe that has, on average, been expanding throughout its history cannot be infinite in the past but must have a past space-time boundary, e.g. the quantum vacuum, which contains quantifiable, measurable energy, and cannot be described as 'nothing', therefore, as 'uncaused' (see \cite{Crai2}).

But, whatever the subtlety of all the arguments used, it remains that to apply human reasoning to the totality of what exists - the universe understood as the set of all things or of all events - even avoiding the paradox of «the set of all sets», is certainly, as Kant had seen, perfectly illegitimate. In any case, it seems difficult to infer from the fact that science, temporarily, marks the step and seems to be bumping against realities momentarily exceeding our understanding, that these realities could be the basis of a consistent extrapolation. It is precisely this kind of transcendental extrapolation that Kant forbade himself, by bringing back the figure of God (who may be for him, however, object of belief), to a simple dressing of the disjunctive syllogism. The only difference with Quantum Physics, is that the alternative, as von Weizäcker already noted (see \cite{Hei}), does not include only 'yes' or 'no' answers, but all the complementary solutions. In this sense, Bohr's theory, even unsatisfactory, will at least have the merit of being a guardrail in front of mystical irrationalism. But the Copenhagen interpretation of Quantum Physics is not, for all that, satisfactory and we must have to wait, before going further, new discoveries. We must have to be creative, for sure, but we must also have to be patient, and not go too fast.

Without doubt, just as rationalism, for Bachelard, became, with the scientific revolutions of the XX$^{th}$ century, a «super-rationalism», so could the expanded naturalism of contemporary science be called, if it did not lead to confusion, a «trans-naturalism». In this sense, the unobservable trans-spatial of Quantum Mechanics could gradually merge into an unknowable trans-natural. Science must stop there. The rest is metaphysics, but if philosophy can legitimately take a look at science (\cite{Bar}), science must remain science.


\begin{thebibliography}{}\addcontentsline{toc}{chapter}{Bibliographie}

\bibitem[Aha 97]{Aha} Aharonov, Y., Vaidman, L., «Protective measurements of two-state vectors» in Cohen, R.S., Horne, M., Stachel, J. (eds.) {\it Potentiality, Entanglement and Passion-at-a-Distance: Quantum Mechanical Studies for Abner Shimony}, vol. Two, 1-8. Springer Science+Business Media, Dordrecht (1997).

\bibitem[Aha 15]{Aha1}  Aharonov, Y, Cohen, E, Shushi T., «Accommodating Retrocausality with Free Will», {\it Arxiv, 1512.06689}, 2015.

\bibitem[Asp 76]{Asp1} Aspect, A., « Proposed experiment to test the nonseparability of Quantum Mechanics », {\it Physical Review D}, vol. 14, no 8,? 15 octobre 1976.

\bibitem[Asp 83]{Asp2} Aspect, A., {\it Trois tests expérimentaux des inégalités de Bell par mesure de corrélation de polarisation de photons}, Thèse de doctorat, Université Paris-Sud, Orsay, 1983.

\bibitem[Asp3 1990]{Asp3} Aspect, A., Grangier, P., «Wave-particle duality: a case study», in Miller, Arthur I. (ed.) {\it Sixty-Two Years of Uncertainty}, 45-59. Plenum Press, New York, 1990.

\bibitem[Bar 10]{Bar} Barreau, A., Parrochia, D. (dir.) {\it Regards philosophiques sur la cosmologie}, Paris, Dunod, 2010.

\bibitem[Boc 14]{Boc} Bôcher, M., «The infinite regions of various geometries», {\it Bull. Amer. Math. Soc.}, 20, no. 4, 185-200, 1914.

\bibitem[Boh 51]{Boh} Bohm, D., {\it Quantum Theory}. Prentice-Hall, Englewood Cliffs, 1951.

\bibitem[Bor 03]{Bor} Borde, A, Guth A and Vilenkin A., Inflationary space-times are incomplete in past directions», {\it Physical Review Letters} 90 (15), 151-301, 2003.

\bibitem[Cat 14]{Cat} Cator, E., Landsman, K., «Constraints on determinism: Bell versus Conway-Kochen», {\it Foundations of Physics}, 44 (7), 781-791, 2014.

\bibitem[Con 06]{Con1} Conway, J., Kochen, S., «The Free Will Theorem», {\it Foundations of Physics} 36 (10), 1441, 2006.

\bibitem[Con 09]{Con2} Conway, J., Kochen, S., «The strong free will theorem», {\it Notices of the AMS}, 56 (2): 226-232, 2009.

\bibitem[Cos 53]{Cos1} Costa de Beauregard, O., «Une réponse à l’argument dirigé par Einstein, Podolsky et Rosen contre l’interprétation bohrienne des phénomènes quantiques», {\it Comptes rendus de l’Académie des Sciences}, tome 236, 1632-1634, 1953.

\bibitem[Cos 79]{Cos3} Costa de Beauregard, O., Le paradoxe d'Einstein (1927) ou d'Einstein-Podolsky-Rosen (1935), Annales de l'Institut Henri Poincaré, 425-433, 1979.

\bibitem[Cos 80]{Cos4} Costa de Beauregard, O., «Cosmos et conscience», in {\it Science et conscience, les deux lectures de l'univers}, Colloque de Cordoue, 1980.

\bibitem[Cos 88]{Cos2}, Costa de Beauregard, O., {\it Le temps déployé, passé - futur - ailleurs}, Monaco, Editions du Rocher, 1988.

\bibitem[Crai 79a]{Crai1} Craig, W. L., {\it The Kalam Cosmological Argument}, Wipf and Stock Publishers, 1979.

\bibitem[Crai 79b]{Crai2} {\it The-existence of god and the beginning of the universe}, Here's Life Publishers, 1979.

\bibitem[Cram 86]{Cram} Cramer, J. G., «The transactional interpretation of Quantum Mechanics», {\it Rev. Modern Phys} 58(July), 647-688, 1986.

\bibitem[Cram 95]{Cram1} Cramer, J., « NASA Goes FTL - Part 2: Cracks in Nature's FTL Armor. "Alternate View" column, Analog Science Fiction and Fact, February 1995.

\bibitem[Cram 14]{Cram2} Cramer, J. G. and Herbert, N., «An Inquiry into the Possibility of Nonlocal Quantum Communication», {\it arXiv:1409.5098}, 2014.

\bibitem[Dur 00]{Dur} Durschmied, E., {\it La logique du grain de sable}, Paris, J.-C. Lattès, 2000.

\bibitem[Ebe 89]{Ebe} Eberhard, Ph. H., Ross, R. R., «Quantum field theory cannot provide faster-than-light communication», {\it Foundations of Physics Letters},2 (2),127-149, 1989.

\bibitem[Eil 13]{Eil} Eilenberg, S. (et alii), «Matter-wave interference of particles selected from a molecular library with masses exceeding 10 000 amu», {\it Physical Chemistry Chemical Physics}, 15(35), July 2013.

\bibitem[Ell 15]{Ell} Ellerman, D., «Why delayed choice experiments do Not imply retrocausality», {\it Quantum Studies: Mathematics and Foundations}, June 2015, Volume 2, Issue 2,183-199, June 2015.

\bibitem[Esp 08]{Esp} Espagnat (d'), B., Saliceti C., {\it Candide et le physicien}, Paris, Fayard, 2008.

\bibitem[Fan 44]{Fan} Fantappie, L., {\it Principi di una teoria unitaria del mondo fisico e biologico}, Roma, Humanitas Nova, 1944.

\bibitem[Fey 65]{Fey} Feynman, R.P., Leighton, R.B., Sands, M., {\it The Feynman Lectures on Physics: Quantum Mechanics, vol. III}, Addison-Wesley, Reading, 1965.

\bibitem[Fit 22]{Fit} Fitzgerald, F. S., «The Curious Case of Benjamin Button», {\it Colliers Magazine}, May1922, 27th, reprinted in F. S. Fitzgerald, {\it  The Curious Case of Benjamin Button and Other Jazz Age Stories}, Penguin Classics, 2008.

\bibitem[Gaa10]{Gaa} Gaasbeek, B., «Demystifying the Delayed Choice Experiments»,  {\it ArXiv 1007.3977}, preprint, 22 July 2010.

\bibitem[Gol 62]{Gol1} Gold, T., «The Arrow of Time», {\it American Journal of Physics}, 30, 403-410, 1962.

\bibitem[Gol 62]{Gol2} Gold, T., «Cosmic processes and the nature of time», in R. Colodny (ed.), {\it Mind and Cosmos}, 311-329, University of Pittsburgh, Pittsburgh, 1967.

\bibitem[Gol 67]{Gol3} Gold, T. (ed.) {\it The Nature of Time}, Ithaca, Cornell University Press, 1967.

\bibitem[Gou 52]{Gou} Goudot, A., {\it Les Quanta et la Vie}, Paris, P.U.F., 1952.

\bibitem[Gra 98]{Gra} Granger, G.-G., {\it L'irrationnel}, Paris, O. Jacob, Paris, 1998.

\bibitem[GHZ 90]{GHZ} Greenberger, D. M., Horne, M. A., Shimony,  Zeilinger, A., «Bell's theorem without inequalities», {\it Am. J. Phys.}, 58 (12), 1131, 1990.

\bibitem[Jac 06]{Jac}, Jacques, V., Wu, E. , Grosshans F., Treussart, F., Grangier, P.,  Aspect, A., «Experimental realization of Wheeler's delayed-choice GedankenExperiment», {\it ArXiv-quant-ph/0610241v1}, 28 oct. 2006.

\bibitem[Har 92]{Har}  Hardy, L., «Quantum mechanics, local realistic theories, and Lorentz-invariant realistic theories», {\it Physical Review Letters}. 68 (20): 2981-2984, 1992.

\bibitem[Har 93]{Har1}  Hardy, L.,  «Nonlocality for two particles without inequalities for almost all entangled states», {\it Physical Review Letters}, 71 (11): 1665-1668, 1993.

\bibitem[Hei 73]{Hei} Heisenberg, W.,  {\it Der Teil und das Ganze – Gespräche im Umkreis der Atomphysik (Physics and Beyond – Encounters and Conversations)}, Munich, 1973.

\bibitem[Jor 48]{Jor} Jordan, P. {\it Die Physik und das Geheimnis des organischen Lebens}, 1941. 6e édition, 1948.

\bibitem[Kim 00]{Kim} Kim, Y.-H., Yu, R., Kulik, S.P., Shih, Y.H., Scully, M.O., «Delayed choice quantum eraser», {\it Phys. Rev. Lett.} 84(1), 2000.

\bibitem[Koc 17]{Koc} Kochen S., «Born's Rule, EPR, and the Free Will Theorem», arXiv.org, 1710.00868. .

\bibitem[Mer 12]{Mer} Mersini-Houghton, L, Vaas R. (ed.), {\it The Arrow of Time}, Berlin, Heidelberg, Springer-Verlag, 2012.

\bibitem[Par 91]{Par1} Parrochia, D., {\it Le Réel}, Paris, Bordas, 1991.

\bibitem[Par 08]{Par} Parrochia, D., {\it La Forme des Crises (logique et épistémologie)}, Seyssel, Champ Vallon, 2008.

\bibitem[Pen  16]{Pen} Penrose, R., {\it The Road to Reality: A Complete Guide to the Laws of the Universe}, Random House, 2016.

\bibitem[Pri 66]{Pri} Price H., {\it Time's Arrow and Archimedes' Point}, Oxford University Press, Oxford, 1966.

\bibitem[Ruy 52]{Ruy} Ruyer, R., {\it Néofinalisme}, Paris, P.U.F., 1952.

\bibitem[Ruy 70]{Ruy1} Ruyer, R., {\it Dieu des religions, Dieu de la science}, Paris, Flammarion, 1970

\bibitem[Scu 91]{Scu} Scully, M.O., Englert, B.-G., Walther, H., «Quantum optical tests of complementarity». {\it Nature} 351, 111-116, 1991.

\bibitem[Sta17 ]{Sta} Stapp, {\it Quantum Theory and Free Will: How Mental Intentions Translate into Bodily Actions}, Springer Verlag, 2017.

\bibitem[Vil 14]{Vil1} Villoresi, P. et alii, «Experimental Satellite Quantum Communications», {ArViv:1406-405Iv1} [quant-ph] 16 juin 2014.

\bibitem[Vil 17]{Vil2} Villoresi, P. et alii, «Extending Wheeler’s delayed-choice experiment to space», {Sci Adv.}, 2017 Oct. 3(10), 2017.

\bibitem[Wer 00]{Wer} Werbos, P. J., Ludmila Dolmatova, L.,  «The Backwards-Time Interpretation of Quantum Mechanics - Revisited With Experiment», {\it arXiv} preprint, 7 August 2000.

\bibitem[Wha 18]{Wha} Wharton, K., «A New Class of Retrocausal Models», {\it Entropy}, 20(6), 2018.

\bibitem[Whe 78]{Whe1} Wheeler, J.A., «The “past” and the “delayed-choice” double-slit experiment» in Marlow, A.R. (ed.) {\it Mathematical Foundations of Quantum Theory}, 9-48. Academic Press, New York, 1978.

\bibitem[Whe 83]{Whe2} Wheeler, J.A., «Law without law, in Wheeler, J.A., Zurek, W.H. (eds.) {\it Quantum Theory and Measurement}, 182-213, Princeton University Press, Princeton, 1983.

\bibitem[Whe 88]{Whe3} Wheeler, J.A., «Hermann Weyl and the unity of knowledge». in Deppert, W., Hübner, K., Oberschelp, A, Weidemann, V. (eds.) {\it Exact Sciences and their Philosophical Foundations Vorträge des Internationalen Hermann-Weyl-Kongresses Kiel 1985} , 469-503, Frankfurt am Main, Verlag Peter Lang, 1988.

\bibitem[Wig 79]{Wig} Wigner, E.P., {\it The problem of measurement. Symmetries and Reflections}, 153-170, Ox Bow Press, Woodbridge, 1979.

\bibitem[Wig  95]{Wig1}  Wigner, E. P., « Remarks on the Mind-Body Question », {\it Philosophical Reflections and Syntheses}, Springer Berlin Heidelberg, coll. « The Collected Works of Eugene Paul Wigner », 1995.

\bibitem[Zei 04]{Zei1} Zeilinger, A., «Why the quantum? “It” from “bit”? A participatory universe? Three far-reaching challenges from John Archibald Wheeler and their relation to experiment», in Barrow, J., Davies, P., Harper, C. (eds.) {\it Science and Ultimate Reality: Quantum Theory, Cosmology, and Complexity}, 201-220, Cambridge University Press, Cambridge, 2004.

\bibitem[Zei 08]{Zei2} Zeilinger, A., «On the interpretation and philosophical foundation of Quantum Mechanics», in: Daub, H. (ed.) {\it Grenzen menschlicher Existenz}, Petersberg, Michael Imhof Verlag, 2008.

\bibitem[Zwi 16]{Zwi} Zwirn, H., «Delayed Choice, Complementarity, Entanglement and Measurement», https://arxiv.org/pdf/1607.02364.pdf, 2016, last version 2017.

\end{thebibliography}
\end{document}